\documentclass{pasj00}
\draft

\begin{document}
\SetRunningHead{Lee et al.}{NEP Wide}
\Received{--}
\Accepted{--}

\title{ North Ecliptic Pole Wide Field Survey of AKARI: Survey Strategy and
Data Characteristics}

\author{
    Hyung Mok Lee \altaffilmark{1},
    Seong Jin Kim \altaffilmark{1},
    Myungshin Im \altaffilmark{1},
   Hideo Matsuhara \altaffilmark{2}
 Shinki Oyabu\altaffilmark{2}
    Takehiko Wada\altaffilmark{2}
    Takao Nakagawa \altaffilmark{2}
	Jongwan Ko\altaffilmark{1}
    Hyun Jin Shim \altaffilmark{1},
    Myung Gyoon Lee\altaffilmark{1}
    Narae Hwang \altaffilmark{1},\
    Toshinobu Takagi\altaffilmark{2}
    Chris Pearson \altaffilmark{2}
    }

\altaffiltext{1}{Astronomy Program, FPRD, Department of Physics and Astronomy,
\\ Seoul National University, Shillim-Dong, Kwanak-Gu, Seoul 151-742,
South Korea} \email{hmlee@snu.ac.kr}
\altaffiltext{2}{Institute of Space and Astronautical
Science, Japan Aerospace Exploration Agency, Yoshinodai 3-1-1, Sagamihara, Kanagawa 229-8510, Japan}


%

\KeyWords{photometry -- infrared: galaxies}

\maketitle

\begin{abstract}
We present the survey strategy and the data characteristics of
the North Ecliptic Pole (NEP) Wide Survey of AKARI.
The survey was carried out for about one year starting from May 2006  with
9 passbands from 2.5 to 24 $\mu$m and the
areal coverage of about 5.8 sq. degrees centered on NEP.
The survey depth reaches to 21.8 AB magnitude
near infrared (NIR) bands, and $\sim$ 18.6 AB maggnitude at the mid infrared
(MIR) bands such as 15 and 18 $\mu$m.
The total number of sources detected in this survey is about 104,000,
with more sources in NIR than in the MIR.
We have cross matched infrared sources with optically identified sources
in CFHT imaging survey which covered about 2 sq. degrees within NEP-Wide survey
region in order to characterize the nature of infrared sources.
The majority of the mid infrared sources at 15 and 18 $\mu$m band are found to be star
forming disk galaxies, with smaller fraction of early type galaxies
and AGNs.  We found that a large fraction (60$\sim$ 80\%) of bright
sources in 9 and 11 $\mu$m is stars while stellar fraction decreases toward fainter
sources.
We present the histograms of the sources
at mid infrared bands at 9, 11, 15 and 18 $\mu$m.
The number of sources per magnitude thus varies as $m^{0.6}$ for longer wavelength sources
while shorter wavelength sources show steeper variation with $m$, where $m$ is the AB
magnitude.
\end{abstract}

\section{INTRODUCTION}\label{sec:introduction}

AKARI (\cite {murakami07}), an infrared space telescope, was launched on February 21, 2006 (UT)
and has successfully carried out its missions including all sky surveys
at mid to far infrared wavelengths and pointed observations at near
to far infared. The `cold' mission lasted until the helium boil out on
August 26, 2007.

In addition to the all sky surveys, AKARI has devoted large amount of
observational time to cover the North Ecliptic Pole (NEP) area
since the NEP (and the South Ecliptic Pole [SEP] as well) is the only
location in the sky with excellent visibility for
polar sun synchronous orbit missions like AKARI. The NEP survey of AKARI is
composed of two programs: Deep (NEP-Deep) and Wide (NEP-Wide) surveys
(\cite{matsu06}). The NEP-Wide observed a large area
($\approx$ 5.8 sq. deg.), while the NEP-Deep covered a smaller (0.38 sq. deg.)
area with integration times longer than the NEP-wide survey by a factor of
3-10.  These two surveys are intended to be mutually complementary.

Both NEP-Wide and NEP-Deep surveys are done with all the filters of
InfraRed Camera (IRC) covering near to mid infrared wavelengths
(see \cite{onaka07}, for the description of this instrument). The IRC wide
bands are designated as N2, N3, N4, S7, S9W, S11, L15, L18W, L24.
The numbers appearing after the roman alphabet represent the approximate
effective wavelengths in units of $\mu$m, while the W's for 9 and 18 $\mu$m represent the fact that
the bandwidths are wider.


Prior to the full scale survey of the NEP, mini surveys were carried out
during the performance verification period in order to check the performance of the
instruments as well as survey strategies. The results of such surveys were
already published (\cite{lee07}, \cite{matsu07}). In addition, AKARI has
regularly observed the `monitor field' near NEP centered at $\alpha =17^h 55^m 24^s$,
$\delta = 66^\circ 37^\prime 32^{\prime\prime}$ with all 9 IRC
bands in order to check the stability of
the instruments during the entire period of the NEP Surveys.
The preliminary analysis of the monitor field data was presented
by \citet{takagi07}.

The NEP-Wide data set has similar but slightly worse (by up to 0.6$^m$)
sensitivity compared to that of the early data of
NEP-Deep as presented by \citet{lee07} who focused on the nature of
11 $\mu$m selected sources covering around 10$^\prime\times 10^\prime$
field of view (FOV).
They found that a majority of sources detected at MIR are
star forming galaxies located at redshift of 0.2 $<z <$ 0.7.
Some of the sources are suspected
to be very red objects at much higher redshift. Since the observed area
was much smaller than the entire NEP-Wide, there were only 72 sources
with 11 $\mu$m magnitude brighter than 18.5. By simple scaling,
we expect to detect around 15,000 sources in the
entire NEP-Wide area with a similar flux limit.

The optical data are very useful in identifying the sources
detected by AKARI since they provide high resolution images.
The optical survey was carried out before the launch of AKARI
covering about 2 sq. deg. inside NEP-Wide survey area
with the CFHT's MegaPrime wide field
camera (see \cite{hwang07}).
We used the CFHT data for the purpose of identification of sources
found in the NEP-Wide survey. The rest of the NEP-Wide area was observed by the
1.5m telescope at Maidanak Observatory in Uzbekistan under very good
seeing conditions ($\lesssim 1^{\prime^\prime}$) with Johnson's B, R, and I filters.
The result of the Maidanak optical survey will be published
elsewhere, but we have utilized both CFHT and Maidanak data
for the purpose of identification of the infrared selected sources.



The purpose of this paper is to present the basic strategies of
the NEP-Wide survey and the properties of the sources found from it.
More detailed analysis of the sources will be postponed to forthcoming papers.
This paper is organized as follows. In the next section,
we describe survey strategy, and the data
reduction process. In \S 3, we present the characteristics, and nature of
the detected sources. We then present the number count band in \S5
and summarize our results in the final section.

\section{NEP-Wide Survey and Data Reduction}\label{sec:data}

\subsection{Survey Strategies}
%

The large area of $\sim$ 5 sq. deg. of the NEP-Wide survey was
required to avoid uncertainty of cosmic variance in the Universe at
$z=0.5\sim 1$, to search for the large-scale fluctuations of the cosmic
near infrared background, and to obtain a large sample of ultra-luminous
infrared galaxies. In order to cover such a large area with small FOV
(10\arcmin $\times $10\arcmin) of IRC for extragalactic study, NEP is the most
suitable area because another high visibility area of South Eclipic Pole (SEP) region
lies very close to the Large Magellanic
Cloud and the visibility is not sufficiently high at any other places.

There were strong constraints on making a large area map using
IRC. First, the FOVs of the MIR-L (long wavelengths of MIR) channel
is separated by approximately
20\arcmin~from that of the NIR and MIR-S (short wavelength of MIR)
channels of AKARI/IRC, although the three channels
take images simultaneously. Second, the position angle of IRC FOVs
depends on the date of the observation because the sun-shield of the
AKARI satellite always has to point to the Sun and thus the focal plane instruments are
aligned along ecliptic coordinates.

\begin{figure}[t]
\begin{center}
    \FigureFile(95mm,95mm){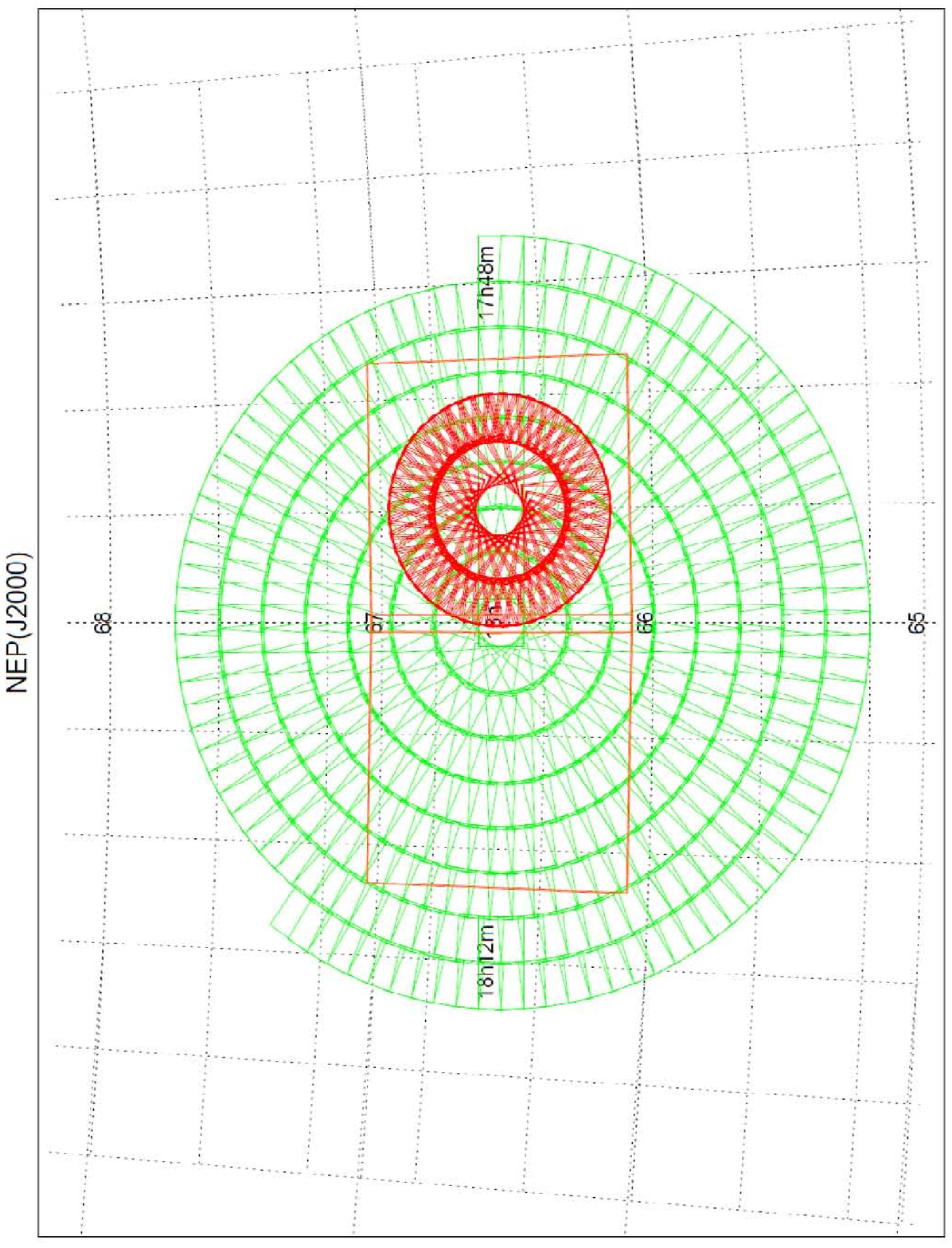}
  \end{center}
   \caption{The survey map of NEP-Wide. The green squares represent the
FOV of individual exposures of IRC. The NEP-Deep exposures are shown
as red squares. In addition, we also show the FOVs of CFHT optical survey.
   }
   \label{survey-map}
\end{figure}

Given these constraints, to maximize the survey efficiency toward the NEP
we planned to make seven concentric circles (and one additional in some
quadrants) centered on NEP with the FOVs of all IRC channels. The survey
was scheduled to be completed within a year
(see Fig. \ref{survey-map}). Also shown in this figure are the NEP-Deep
exposure map indicated by red boxes as well as the optical survey area of CFHT as yellow
boxes.
For the observational redundancy, an observation on a circle was planned
to be shifted with half of FOVs of neighboring observations, so that any
area would be observed at least twice.
Each pointing observation was done by the `IRC03' template.
This observing mode was designed for general purpose imaging observations that
take images with three filters in a pointed observation. For each filter
two imaging observations are made with dithering operations.
The detailed observational procedure for each Astronomical Observation
Template (AOT) is described in AKARI IRC Data User Manual (\cite{lorente07}).

The observations were carried out as planned as long as the schedule permits.
The NEP-wide survey was completed with 446 pointing observations.

\subsection{Pre-Processing}

The observed data were processed using IRC data reduction pipeline
version 071017.  The pipeline subtracts the dark current, normalizes
the signal, corrects for the linearity, and applies
the flat field corrections.  The cosmic rays are removed in the pipeline
for the MIR images (S7, S9W, S11, L15 and L18W) during the image stacking
process because these images are taken
with multiple exposures per pointing. However, the number of exposures for
NIR bands was smaller than that of MIR bands, and therefore the cosmic
rays have not been completely removed during the stacking of multiply exposed
images. Especially, there are only two frames to be stacked in most of the N4
data so that the majority of cosmic rays remain unremoved.
In such cases, we have applied a program L.A.cosmic (\cite{dokkum01})
which employs Laplacian edge detection algorithm.
The cosmic ray removal for N4 band image data heavily relied on L.A.cosmic task which
often makes somewhat erroneous corrections for the cosmic ray effects.
Therefore we should take the N4 band photometric data with more care
than those of other wavelengths.
We applied L.A.cosmic for N2 and N3 as well. For these data, large fraction of cosmic rays can be
removed during the stacking process.

%



Finally the astrometric information
is added. The astrometric solution can be found nearly automatically for
N2, N3, N4, S7 and S9W images by comparing with the 2MASS database.
However, the mid-infrared band frames for S11 to
L24 contain relatively smaller numbers of sources so that often automatic
astrometric solution cannot be derived.
In such cases, we have used imaging data with astrometric solutions at
shorter wavelengths in order to obtain the accurate positional information for
next longer wavelength images (i.e., S9W for S11, etc).
This procedure provides reasonable results up to L18W images for most exposures.
However, the procedure was successful for only 239 frames out of 446 for L24 band due to
a small number of identifiable sources. Because of such a poor
astrometric calibration we do not present the analysis of the L24 band
data in this study. We will report the result for L24 band in forthcoming
papers.

After the astrometric solutions are settled for individual pointing data, we
mosaicked them to produce a single master image of entire survey region for each
passband using SWarp software package by E. Bertin at Terafix
(see {\tt http://terapix.iap.fr/IMG/pdf/swarp.pdf}
for the detailed description of this software).


The procedure to obtain photometric data is described in
next subsection. Note that magnitude system referred henceforth is AB magnitude.

\subsection{Point Spread Function}

In order to check the quality of the mosaic images, we have measured
the FWHM of bright point sources in mosaic images and listed the results in Table \ref{PSF},
together with single frame FWHMs taken from IRC Data User Manual (\cite{lorente07}).
In the case of NIR bands, the mosaic images have about 30\% larger
FWHM than single frame values while the degradations
in S and L bands are much smaller. The degradation in NIR band is mostly due to the elongated shape
of PSFs in NIR bands.

\begin{table}[t]
\caption{Full Width at Half Maximum (FWHM) in units of pixel (and in arcseconds in the parenthesis)
of the Point Spread Function}
\label{PSF}
\begin{center}
\begin{tabular}{ c c c c c c c c c}
\hline & N2 & N3 & N4 & S7 & S9W & S11 & L15 & L18W\\
\hline \hline
Single Frame&
2.9 (4.2 $^{\prime\prime}$) & 2.9 (4.2 $^{\prime\prime}$)& 2.9(4.2 $^{\prime\prime}$)& 2.2
(5.1 $^{\prime\prime}$)& 2.4 (5.1 $^{\prime\prime}$)& 2.4 (5.1 $^{\prime\prime}$)& 2.3 (5.6$^{\prime\prime}$)
& 2.3(5.6$^{\prime\prime}$)\\
Mosaic & 3.8 (5.5$^{\prime\prime}$)& 4.1 (6.0$^{\prime\prime}$)& 4.1 (6.0$^{\prime\prime}$)&2.5 (5.9$^{\prime\prime}$)
 &2.8 (6.6$^{\prime\prime}$) & 2.5 (5.9$^{\prime\prime}$)& 2.5 (6.2 $^{\prime\prime}$) & 2.5 (6.2$^{\prime\prime}$)\\
\hline
\end{tabular}
\end{center}
\end{table}

\subsection{Coverage and Noise Maps}
\begin{figure}[t]
\begin{center}
    \FigureFile(90mm,100mm){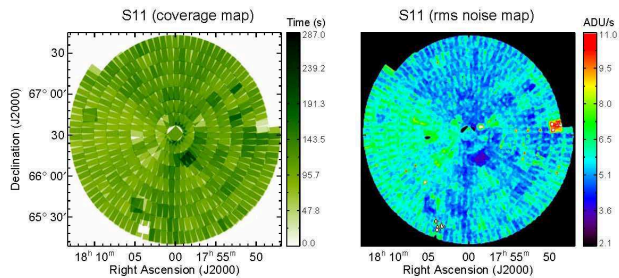}
  \end{center}
   \caption{The coverage map which is analogous to the map of integration
times, of NEP-Wide survey at S11 band (left panel) and noise map (right panel). }
   \label{s11_coverage}
\end{figure}

Since the survey was carried out with multiple, but non-uniform number of pointings,
the integration time varies over the survey region. The number of pointings ranges
from 1 to 3 and most regions are covered twice. The integration time per pointing
are listed in Table \ref{int_time}. The actual integration time for given
line of sight can be obtained by multiplying the single pointing integration time in
Table \ref{int_time} with the number of pointings of that direction. Single
pointing integration varies from 89 (N4) to 147 seconds (S7, S9W, L15 and L18W).
The noise levels depends on the total integration time.
Fig. \ref{s11_coverage} shows the coverage map of S11, as an example,
which corresponds to the distribution of integration time over the survey
region.

\begin{table}[t]
\caption{The integration time per exposure, number of exposures per pointing in that direction,
and integration time
per pointing. The integration time units are in seconds}
\label{int_time}
\begin{center}
\begin{tabular}{ c c c c c c c c c}
\hline & N2 & N3 & N4 & S7 & S9W & S11 & L15 & L18W\\
\hline \hline
per exposure&44.4 & 44.4 & 44.4 & 16.36 & 16.36 & 16.36 & 16.36 & 16.36 \\
Number of Exposures & 3 & 3 & 2 & 9 & 9 & 6 & 9 & 9 \\
per pointing & 133.2 & 133.2 & 88.8 & 147.24 & 147.24 & 98.16 & 147.24 & 147.24 \\
\hline
\end{tabular}
\end{center}
\end{table}

We also estimated the noise levels throughout the survey area by measuring the
{\it rms} fluctuations, using
16$\times$16  nearby pixels centered on each pixel. The poorly covered
regions are expected to have large {\it rms}
fluctuations while well sampled regions have small fluctuations.
Most regions have {\it rms} fluctuations less than 6 ADU/s which corresponds
to point source detection limit (5-$\sigma$) of 19-th magnitude. The very noisy
region shown in the right hand side of the {\it rms} map was exposed
only once, and therefore we were not able to properly remove
cosmic rays.

We found that a sector at RA=$17^h 50^m$ and Dec=$66^\circ 35^\prime$
in the Fig. \ref{s11_coverage}  has systematically
higher noise levels. The AKARI has suffered from the Earth shine problem during
the summer of 2007 when more stray light entered the focal plane.
The data obtained during this period show systematically higher
background level in all filter bands. The higher background level means higher
poisson noise, and such an effect is shown in the {\it rms} map of Fig.
\ref{s11_coverage}.

\begin{figure}[t]
\begin{center}
    \FigureFile(90mm,100mm){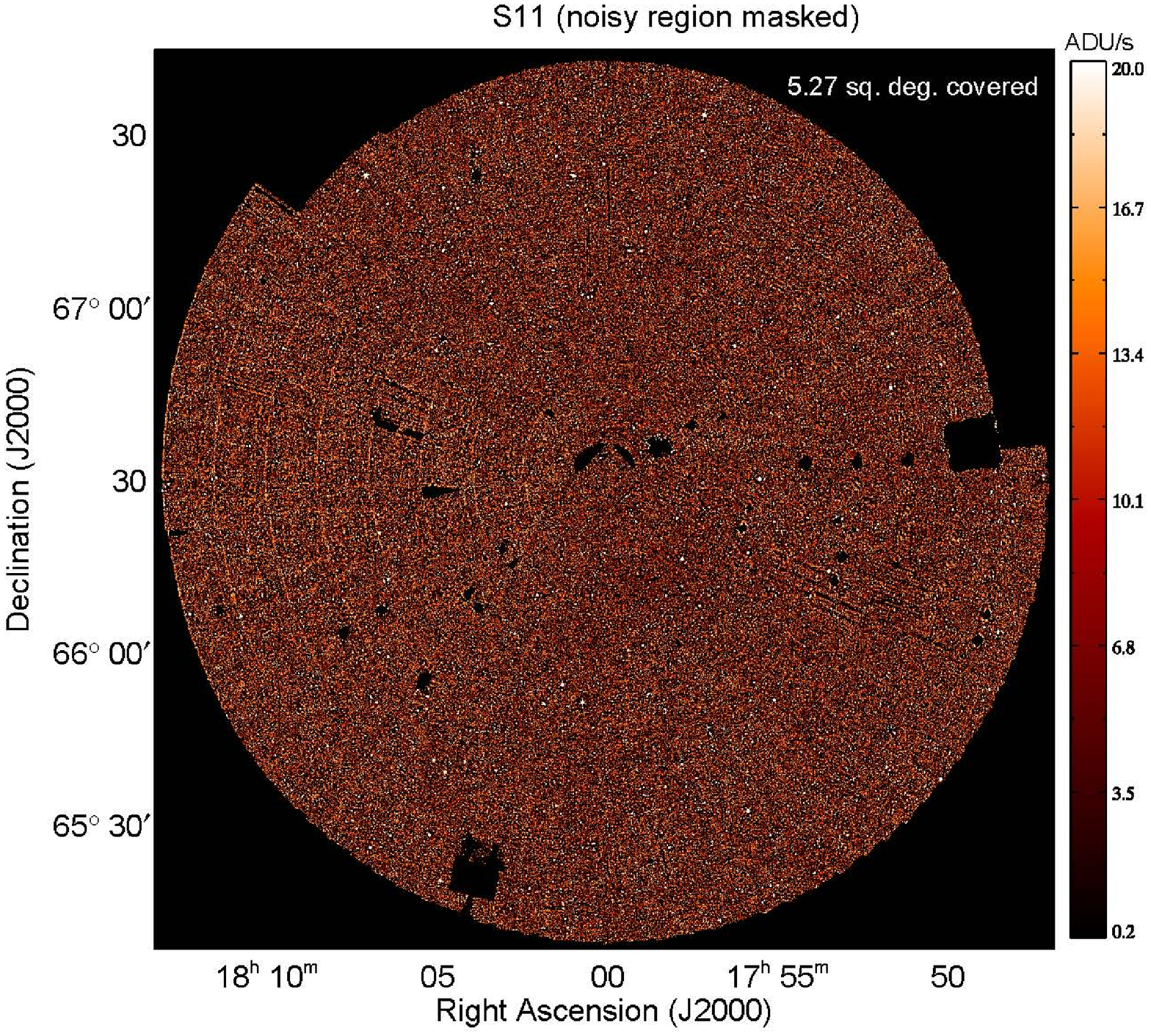}
  \end{center}
   \caption{The map of the survey area excluding the high noise region.}
   \label{s11_lownoise}
\end{figure}

After making the noise map, we produced a low-noise maps
excluding high noise regions in order to avoid too many false detections.
In Fig. \ref{s11_lownoise},
we show an example of S11 band image
with {\it rms} fluctuations smaller than 6 ADU/s.

The photometry was done using this low-noise map. The total area of S11 map was 5.35 sq. deg. and
the area excluded due
to high noise level is around 0.12 sq. deg. leaving 5.27 sq.  deg. of low-noise region.


\subsection{Photometry}

We used
SExtractor (\cite{bertin96}) to detect the sources in the indivisual
image of each band.
We have confirmed a sources if it has more than five contiguous pixels
above 3 times of {\it rms} fluctuations of the sky. The intensity
of each source was measured within the isophotal
aperture that satisfies the criteria for the source detection.
The photometry was done using
SExtractor in single mode.


The detection limits depend on the noise levels which vary from
place to place. Since the low-noise map shown in Fig. \ref{s11_lownoise} covers
the region with {\it rms} fluctuation less than 6 ADU/s, the
5-$\sigma$ detection limit should be better than 19-th magnitude at
11 $\mu$m. The resulting average 5-$\sigma$ detection limits for
eight photometric
bands are listed in Table \ref{det_limit}, together with the number of sources detected over
3-$\sigma$ detection threshold.


\begin{table}[t]
\caption{5$\sigma$ detection limits and number of sources detected over 3$\sigma$ threshold for 8 filter bands }
\label{det_limit}
\begin{center}
\begin{tabular}{ c c c c c c c c c}
\hline & N2 & N3 & N4 & S7 & S9W & S11 & L15 & L18W\\
\hline \hline
$5\sigma$ detection limit (AB Mag.)& 21.5& 21.8& 21.7& 19.5& 19.4&18.9 &18.5 & 18.7 \\
Number of detected sources& 92,731 & 104,014& 98,704 &14,545 &18,366 & 15,561& 11,713& 14,087\\
\hline

\end{tabular}
\end{center}
\end{table}

The number of sources detected
in individual filters vary significantly. There are much more sources in near
infrared than mid infrared.  For example, there are about 104,000 sources in
N3 band while about 14,500 sources are detected in S7.

\subsection{Astrometric Accuracy}

\begin{figure}[t]
\begin{center}
    \FigureFile(90mm,100mm){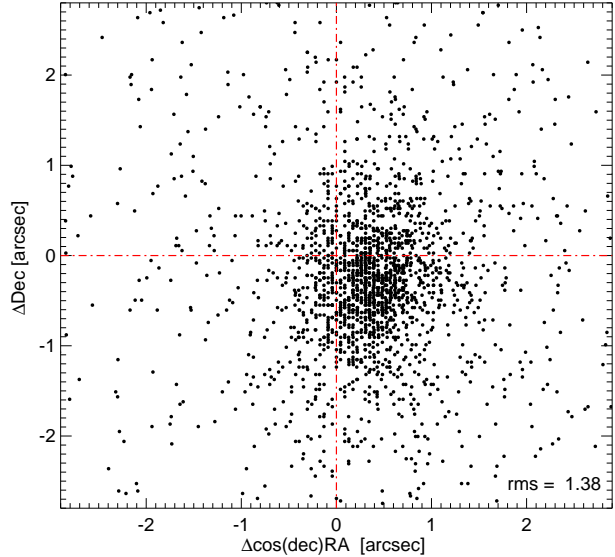}
  \end{center}
   \caption{The scatter diagram of astronometric positions between the
NEP-Wide data and the CFHT. We found that there is an offset between CFHT
and NEP-Wide astrometric solutions of about 0.39 arcsecond. The {\it rms}
deviations shown in the figure is
about 1.38 arcsecond which is about 30\% of the FWHM of the single frame PSF
of the NIR bands of AKARI.}
   \label{astrometry}
\end{figure}

In order to check the accuracy of our astrometric solutions, we
compared the positional data of AKARI's point sources that are brighter
than 19-th magnitude in N2 and 18-th magnitude in S11, with the counterpart sources in CFHT data.
For this purposes, we matched our source list
with that of the CFHT data base by \citet{hwang07}.
The matching radius was chosen to be 3.5 arcseconds
and the number of matched sources was about 2000.
The astrometric comparison is shown in Fig. \ref{astrometry}.
We found that there is an offset of $\Delta (\cos\delta \times \alpha) =0.33^{\prime\prime}$
and $\Delta\delta = -0.2^{\prime\prime}$ between CFHT and NEP-Wide
astrometric solutions (AKARI-CFHT).
This offset is known to exist when CFHT data were compared with
Guide Star Catalogue (GSC, \cite{jenkner90}).
The {\it rms} deviations shown in the figure is
about 1.38$^{\prime\prime}$ which is about 30\% of the FWHM of the single frame PSF
of the NIR bands.

\subsection{Photometric Consistency}
In order to check the accuracy of our photometry, we compared our catalogue with that of
NEP-Deep field by Wada et al. (2008) in Fig. 5.
The NEP-Deep field lies within the
NEP-Wide survey area as shown in Fig. 1.
We found that the scatter is becomes larger at fainter magnitudes, and there exist
small but non-negligible difference (order of 0.02 - 0.06 magnitudes) between NEP-Wide and
NEP-Deep in the sense that NEP-Wide is slightly fainter When checked with the Monitor Field
data by Takagi et al. (2008), we found opposite tendency, although the difference is also
very small.
The systematic difference may have resulted from the different photometry parameters.
The small differences of $\lesssim 0.06$ magnitude
would not affect our conclusions of the present work of characterization of the data and source.

\begin{figure}[t]
\begin{center}
    \FigureFile(90mm,100mm){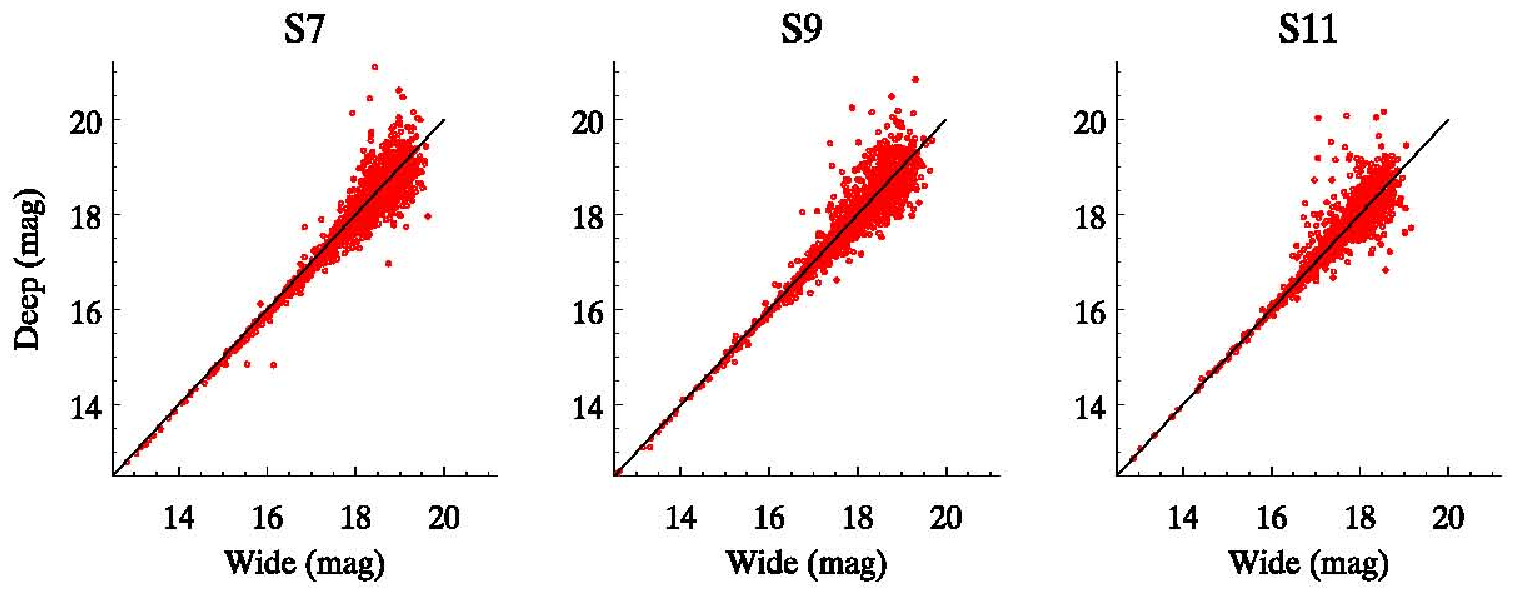}
    \FigureFile(90mm,100mm){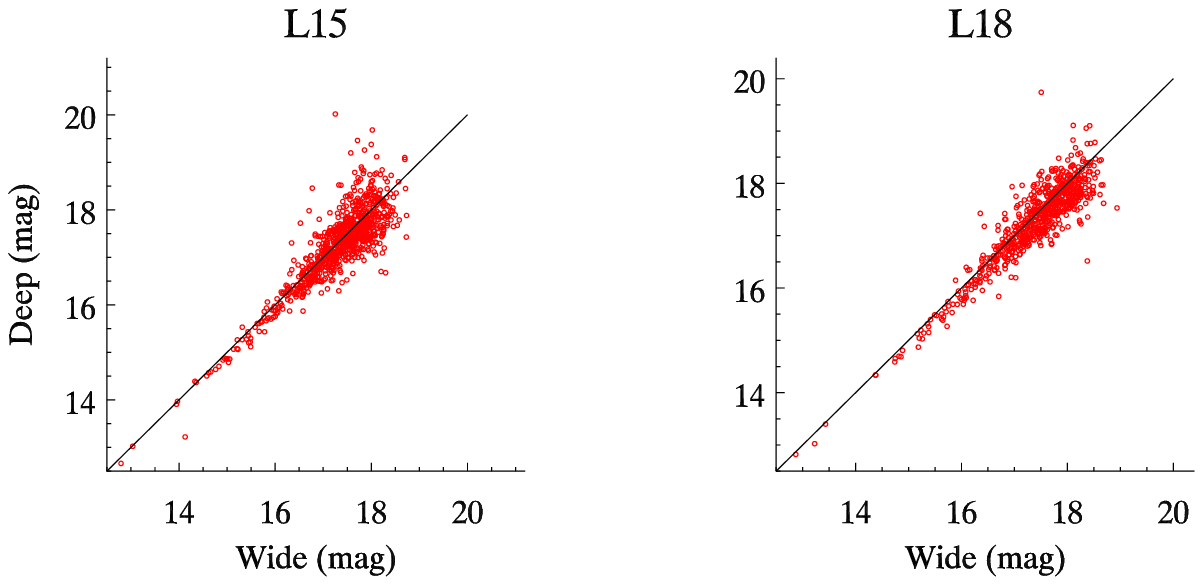}
  \end{center}
   \caption{The comparisons of the photometric magnitudes between the NEP-Deep
data by Wada et al. (2008) and the NEP-Wide data of the present work for N-
and S-bands. There are small but non-negligible
systematic difference between these two data sets. NEP-Wide is fainter by about 0.02 $\sim$ 0.06 magnitude.}
   \label{comparison}
\end{figure}

\section{Nature of the Detected Sources}\label{sec:nature}

The sources detected in our survey ranges from various Galactic objects to
distant galaxies and quasars. Although it is not easy to identify the nature
of all the sources using the AKARI data alone, we try to categorize them based
on our photometric data and optical survey data. It is worth noting that
the previous study by \citet{lee07} for the early data of NEP-deep showed
that the majority of the S11 flux limited sample are
star forming galaxies while less than 20\% of the detected sources are
found to be stars. Since the NEP-wide survey has a sensitivity similar to that
of the sample discussed in \citet{lee07}, we also expect a large fraction of the
mid-infrared selected sample to be star forming galaxies.

\begin{figure}[t]
\begin{center}
    \FigureFile(90mm,100mm){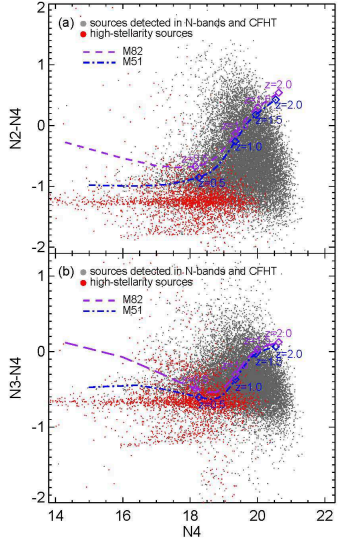}
  \end{center}
 \caption{The N4 versus N2-N4, and N4 versus N3-N4 CM diagrams sources that are
   matched with CFHT optical sources in the CFHT survey field (\cite{hwang07})
    with optical counter parts within 3.5 arcseconds, and S11
magnitude brighter than 18.5.
The red dots represent the sources whose $r^\prime$
stellarity greater than 0.8 and $r^\prime$ magnitudes brighter than 19.
Also shown here are the tracks of star forming galaxies as represented by
template SEDs of M51 and M82 by \citet{silva98}.}
   \label{cm1}
\end{figure}

In order to distinguish objects with different natures, we generated
color-magnitude (CM) and color-color (CC) diagrams for the sources
that are matched with the CFHT point sources. In Fig. \ref{cm1}, we show
the CM diagrams of NEP-Wide sources with S11 brighter than 18.5 magnitude.
The red dots represent the sources whose  stellarity parameters ($sgc$) are
greater than 0.8  in the CFHT $r^\prime$-band image and $r^\prime$ magnitude brighter than
19. These sources are thought to be mostly stars, and we have
confirmed that the optical-IR spectral energy distribution (SED) of these objects
indeed closely follow that of black-body, assuring that these are stars.
As expected from the black body nature of stellar SEDs, the
star-like sources have NIR
colors lying in rather narrow ranges in N2-N4 and N3-N4.
The number of sources with $S11<18.5$, $sgc>0.8$ and $r^\prime <19$ are 780
while total S11 flux limited sample within the CFHT survey field is 3800.
This means that the stellar fraction is about 20\% for a sample with $S11<18.5$.
The $r^\prime$ brightness cut was applied to avoid possible contamination by
faint galaxies.
This can be compared with the previous study by \citet{lee07} who found that
the stellar fraction is about 18\% of the 72 S11 flux limited sample after
careful examination of optical to MIR SEDs as well as the optical
images. Thus our criteria of $sgc>0.8$ for bright optical sources can detect
all the stars and could possibly include some distant compact galaxies.

Also shown in these CM diagrams are the tracks of
star forming galaxies represented by M82 (violet lines) and M51 (blue lines)
SEDs. We have taken the template SEDs of these star forming galaxies from
\citet{silva98}. These galaxies are taken as typical examples of
different star formation rates (SFR): M51 for low SFR and M82 for
moderate SFR galaxies. Although the rest frame near infrared color depends on
the star formation rate significantly, the tracks converge into nearly identical lines at
redshifts greater than 0.5. Since we assumed the luminosity to be $L_*$, the
tracks of galaxies with different luminosities would slide horizontally from
the lines shown in this figure so that the sources at lower right corner
can also be explained by fainter star forming galaxies lying at
moderate redshifts ($z \lesssim 1.0$).

\begin{figure}[t]
\begin{center}
    \FigureFile(90mm,100mm){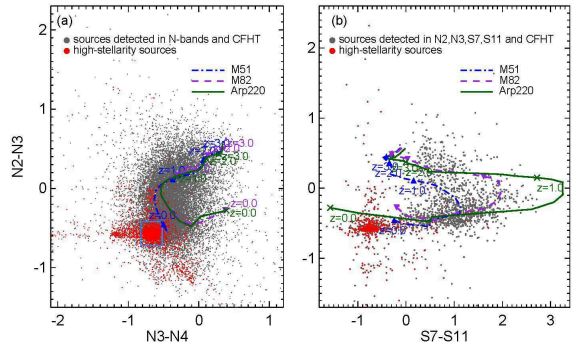}
  \end{center}
   \caption{The N2-N3 versus N3-N4 and N2-N3 versus S7-S11 CC diagrams
for the same sources shown
   in Fig. \ref{cm1}. The symbols are the same as those in Fig. \ref{cm1}
}
   \label{cc1}
\end{figure}

In Fig. \ref{cc1}, we show near and mid infrared CC diagrams.
The red dots are the `stellar sources' as defined by the stellarity
parameter for bright optical sources. As expected from the narrow ranges in
N2-N3 and N3-N4 colors, majority of the stellar sources lie in a small
blue box shown in the upper panel. The tracks of star forming galaxies
are also shown in these CC diagrams. Here we have added Arp 220 which is a
typical example of starburst galaxy. While the tracks are
highly degenerated in the NIR CC diagram in the upper
panel, the MIR-NIR CC diagram shown in the lower panel is more sensitive
to the star formation rates as well as the redshifts.  For the galaxies with
very high star formation rates (e.g., Arp 220), rest frame S7-S11 color is very blue because of
very strong 7.7 $\mu$m PAH feature. However as this PAH feature redshifts to
11 $\mu$m band, the tendency of the color is reversed. As seen in the lower panel
of the Fig. \ref{cc1}, redder part of the S7-S11 color is occupied
by the galaxies with higher star formation rate located at ($z \lesssim 1.0$).

\begin{figure}[t]
\begin{center}
    \FigureFile(90mm,100mm){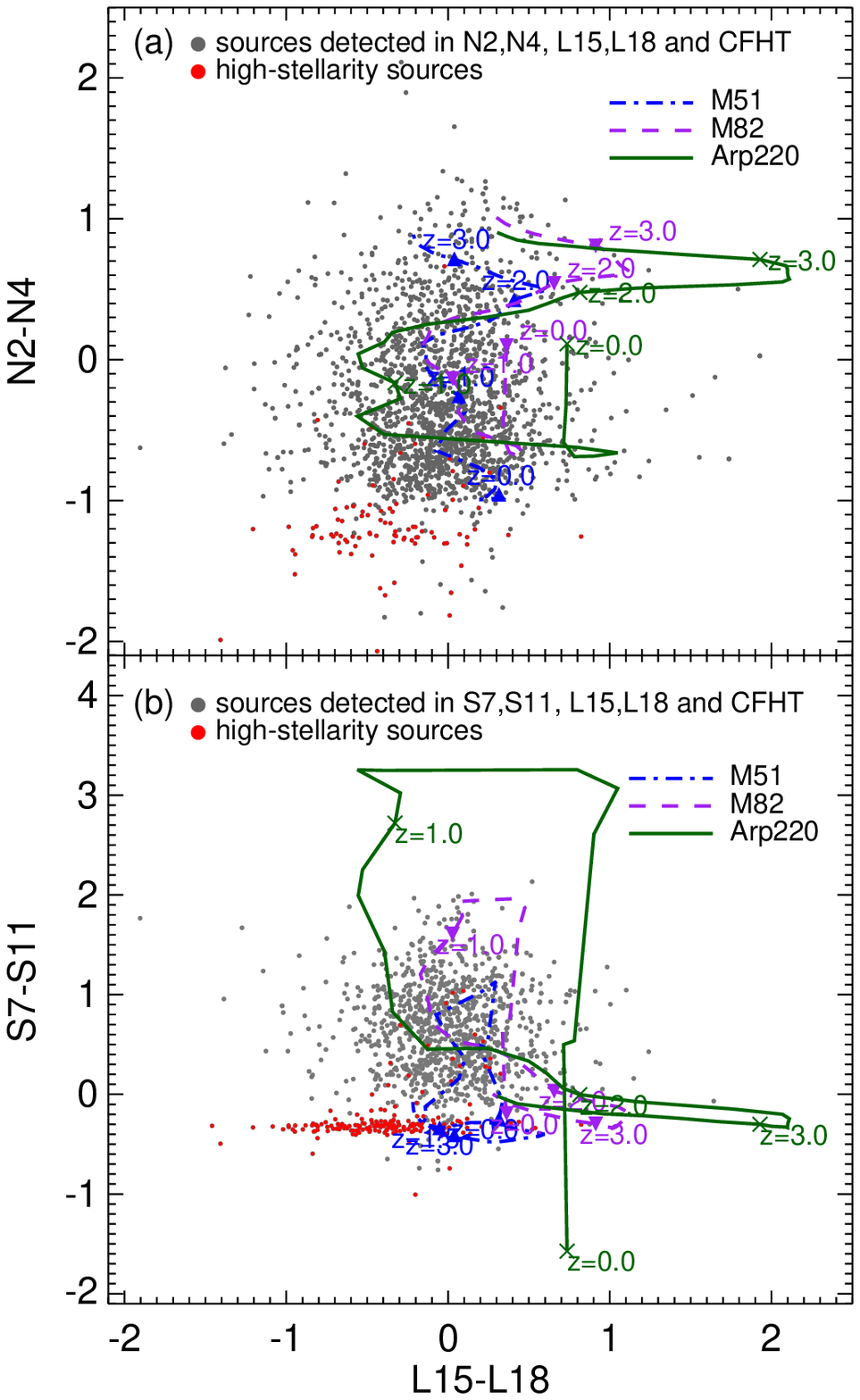}
  \end{center}
   \caption{The S7-S11 versus L15-L18W CC diagrams for the same sources shown
   in Fig. \ref{cc1}.}
   \label{cc2}
\end{figure}
In Fig. \ref{cc2}, we  have shown two more CC diagrams with MIR (S7-S11 or L15-L18W)
and NIR colors (N2-N4). The upper panel shows the CC diagram of N2-N4 versus L15-L18W
while the lower panel show the CC diagram of S7-S11 versus L15-L18W. The
tracks of three types of star forming galaxies represented by M51, M82 and Arp 220.

Again we find that the NEP sources occupy large area in these CC diagrams.
When compared with
the tracks of various types of star forming galaxies, we find that majority of
NEP sources are star forming galaxies within moderate redshifts. However, we also
notice that many sources fall on some parts of CC plane which cannot be occupied by
star forming galaxies such as lower left part and lower right of
the N2-N4 versus L15-L18W CC diagram. Also the rare objects with very red N2-N4 color are likely to
be AGNs (\cite{lee07}). The range of L15-L18W color range of the NEP-Wide sources
appears to be much wider than typical star forming galaxies.
Unlike S7-S11, L15-L18W color does not seem to be an indicator of SFR.
Clearly more careful study is needed
to identify the nature of the sources found in the NEP-Wide survey.

\begin{figure}[t]
\begin{center}
    \FigureFile(90mm,100mm){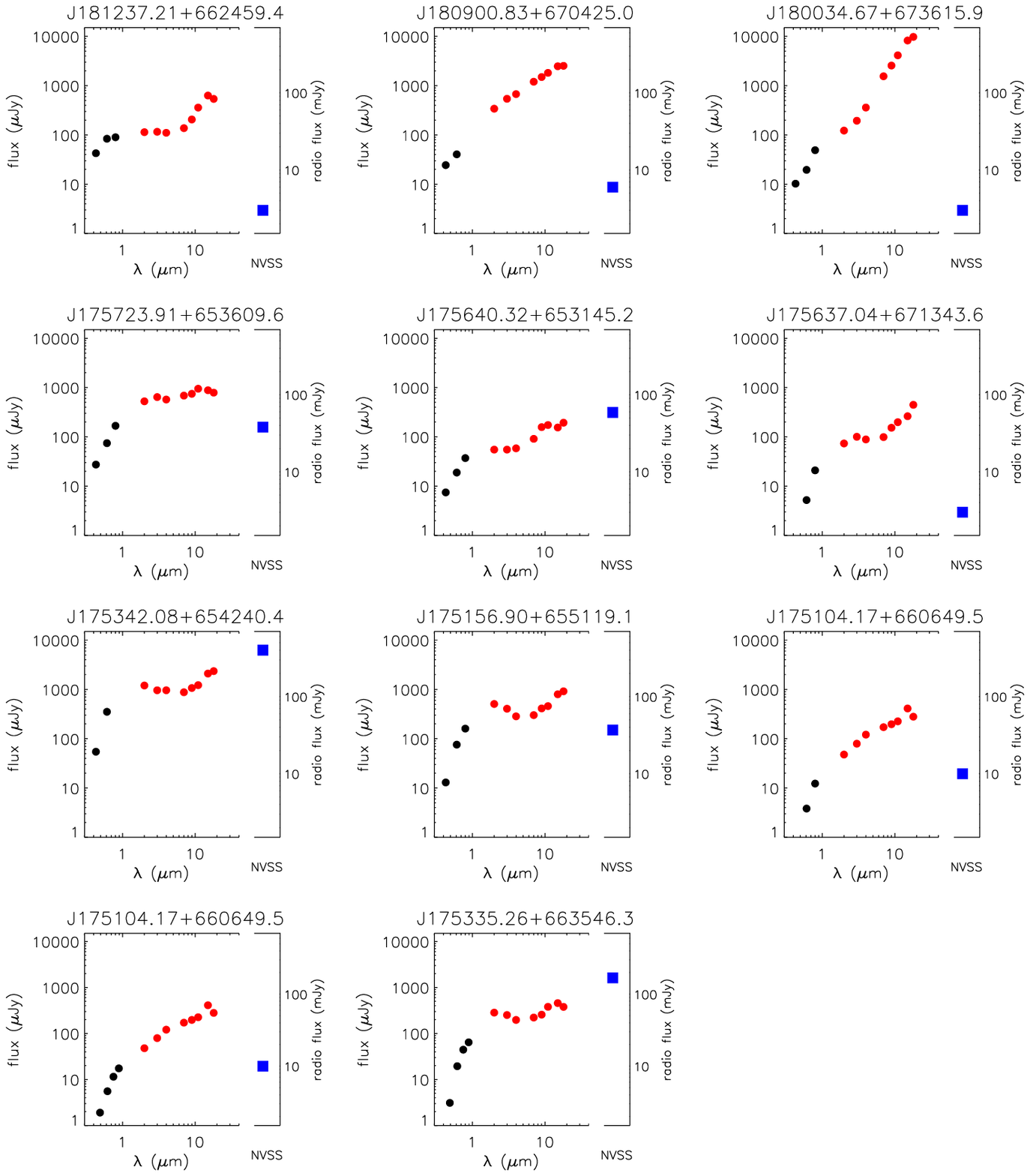}
  \end{center}
   \caption{Optical-MIR SED plots for the radio sources with MIR and optical
   counterparts. The radio fluxes at 1.4 GHz are indicated as blue dots.
   Four objects show clear power-law behavior of the AGNs.
   There are also several sources which show NIR bumps indicating the
   presence of late type stars. These objects need more careful examination, but some of them
   could also be AGNs.
}
   \label{radio}
\end{figure}

We attempted to identify MIR sources among the radio emitting objects from the
NRAO/VLA Sky Survey (NVSS) data (Condon, et al. 1998). Out of about 260 radio sources brighter than 2.5
mJy at 1.4 GHz (21 cm) in NEP-Wide area, we found 30 sources that match with the MIR source
at 15 $\mu$m. In Fig. \ref{radio} we show the optical-MIR SED plots of the 11 sources which
have optical counterparts in CFHT or Maidanak data. The radio flux is shown as a blue dot
in the righthand side of each plot. Although there appears no clear correlation
between MIR and radio fluxes, we found that the large fraction of the radio sources
with MIR counterparts are suspected to be AGNs. For example, 4 out of 11 ($\sim 36$\%) sources show
power-law SEDs (J180900.83+670425.0, J180034.67+673615.9, J175104.17+660649.5 and J175204.17+660649.5),
indicating they are indeed AGNs. This can be compared with the finding by Lee et al. (2007) that
only 6\% of the MIR selected sources
are AGNs. The remaining 7 sources which show NIR bumps
need more careful examination since some of them could also be
AGNs having large amount of late type stars.

\begin{figure}[t]
\begin{center}
    \FigureFile(90mm,100mm){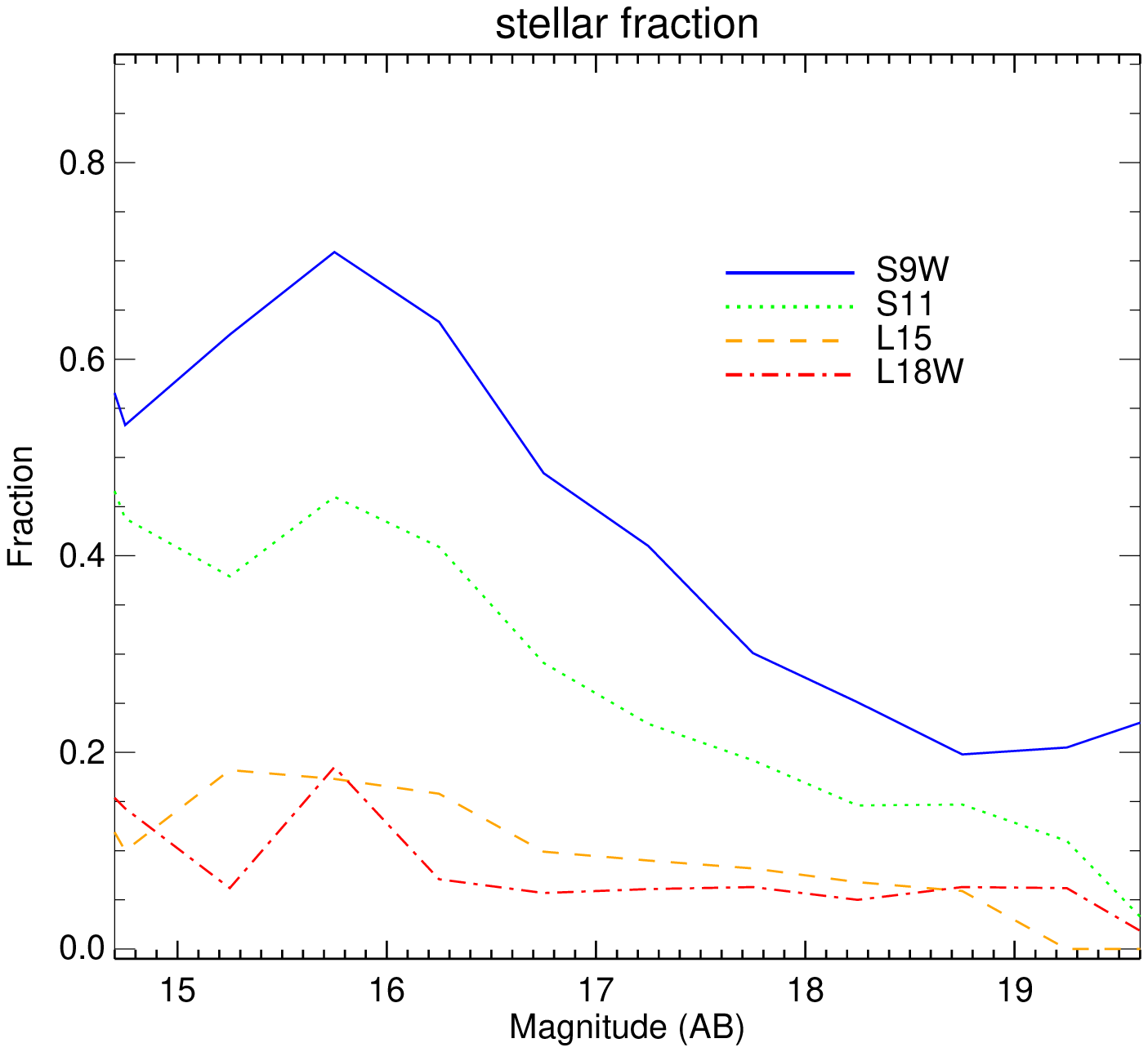}
  \end{center}
   \caption{The fraction of stellar sources in mid-infrared band sources
as a function of the MIR magnitudes.
The sources in S9W and S11 bands contain many stars, but
the stellar fraction becomes much smaller for L15 and L18W bands. Also,
the stellar fraction decreases toward the fainter ends in all bands.
}
   \label{star_frac}
\end{figure}

It is important to distinguish between stellar and extended sources
for further analysis. As we saw in the comparison with the NEP-Early
data, almost all the sources with stellarity $sgc>0.8$ are likely to be
stars. We estimated the fraction of stars among the mid-indrared
sources by using the CFHT data. We identified the sources as stars
when the mid-infrared sources have optical counterparts in the
CFHT field.
The `stellar fractions' determined in this way in S9W, S11, L15 and
L18W band sources are
shown in Fig. \ref{star_frac} as a function of magnitude. The size
of the magnitude bin was 0.5 mag.
In S9W and S11, the stellar fraction
is very large for bright sources and decreases rapidly toward the
fainter ends. However, the stellar fraction of the sources detected
in L15 and L18W are much smaller than that of S11, and it decreases
rather slowly toward the fainter parts. There is virtually no sources with high
stellarity among faint sources.

\section{Flux Distribution in Mid Infrared}
One of the powerful ways for the study of the evolution of galaxies is to
examine the distribution of sources as a function of flux. We know
that the source counts flatten at faint flux but the magnitude of
the counts provides a strong constraint on source count models by
constraining the local
luminosity function such as $L_*$ and $\phi$ of the
Schechter luminosity function since different assumptions on these
parameters can produce very different normalizations.
For observed source counts, in particularly at 15$\mu$m we will be able to
confirm which of the ELAIS fields (ELAIS N or S) is giving us the right
picture since the ELAIS S field has a much lower normalization
(Gruppioni et al. 2002).
In order to do this, we need to
accurately separate the galaxies and stars, estimate the incompleteness, and
assess effects of saturation for bright sources, among others.
Also, a combination of
NEP-Deep and NEP-wide data is necessary for more accurate source count over
wide range of fluxes.

\begin{figure}[t]
\begin{center}
    \FigureFile(90mm,100mm){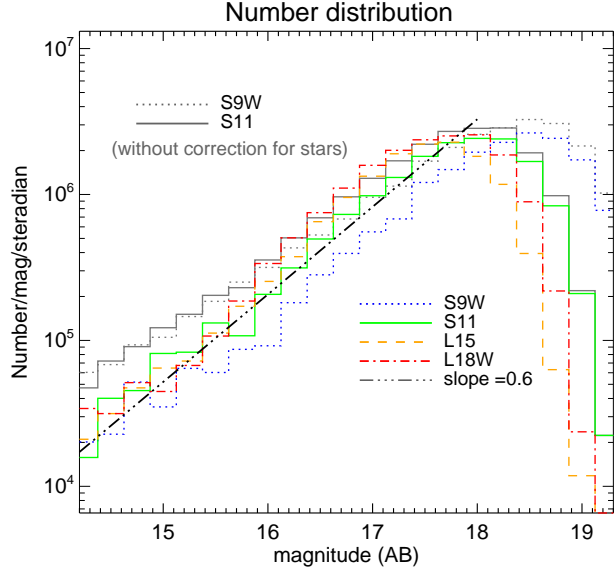}
  \end{center}
   \caption{The histogram of MIR sources in NEP-Wide survey for
   S9W, S11, L15 and L18W bands. The slopes for L15 and L18W at brighter end are
   close to 0.6, as expected for uniform distribution in Euclidean space while
   S9W and S11 slopes are slightly steeper. The count plots after correcting for the
   stars (i.e., applying the fraction of non-stellar objects) are also shown for
   the cases of S9W and S11 since the stellar fractions in these bands
are significant
   and have strong dependence on magnitude. The corrected counts are consistent with
   the uniform source distribution in flat universe.
}
   \label{histogram}
\end{figure}

Therefore we leave such tasks for subsequent works.
Instead of showing differential count in a more accurate way, we simply show the
histograms of the NEP-wide sources detected in S9W, S11, L15 and L18W in
Fig. \ref{histogram}. The expected
slope for uniform source distribution in Euclidean geometry is also shown in
this histograms.

For the case of S9W and S11, we have shown both the raw count without taking out
stars, as well as the count for the non-stellar objects (i.e., galaxies) by
statistically correcting for stellar sources using the stellar fraction of
Fig. \ref{star_frac}. Since the stellar fractions in L15 and L18W are relatively
small, the correction does not make any significantly different
histogram.

We find that the slopes of the number distribution for L15 and L18W
sources are close to that of uniform distribution in Euclidean space
(i.e., 0.6) while the
raw data of S9W and S11 show slightly less steep distribution.
However, the flux distribution
for non-stellar sources after correcting for stellar fraction becomes
more consistent with
the uniform source distribution, although the actual slopes in these
passbands are still slightly
flatter than L15 and L18W bands.

Such a difference might be attributed by strong positive K-corrections from
the 7.7 $\mu$m feature (in the S9W band) and 11.3 $\mu$m (in the S11 band) PAH
features. These features will be local at the brightest magnitudes and the
decrease in slope seen to fainter magnitudes could be caused by these
emission features leaving the pass bands resulting a decrease in flux. This
would also lead to a non-Euclidean slope at bright magnitudes in the short
wavelength counts.
The L15 and L18W band counts generally sample the higher redshift Universe
and are not so affected by local contributions.

Note that a similar effect was observed in the IRAC 8 $\mu$m band which
straddled the 7.7 $\mu$m emission feature.

\section{Summary}
The NEP-Wide survey of AKARI covered about 5.8 sq. deg. region centered on
NEP. The survey data consisting of 446 pointing observations at 8 band filters from 2 to 18 $\mu$m
are reduced. Total number of sources detected at least in one filter band reaches
about 104,000.
However, the numbers of sources at MIR passbands are
significantly smaller than those at NIR.
The 5-$\sigma$ detection limits are typically about 21.5$^m$ 2-5 $\mu$m bands,
18.5 to 19.5$^m$ for
7 to 18 $\mu$m bands. Our data will be complemented by the NEP-Deep Survey
data which are deeper than
NEP-Wide with smaller areal coverage.

We used the CFHT optical data which covered 2 sq. deg. within the NEP-Wide survey
area to identify the nature of the objects. In particular, we have used
the stellarity
parameter of SExtractor to distinguish between stars and non-stellar objects which
are external galaxies.

The majority of mid infrared sources are thought to be star forming galaxies at
redshift $\lesssim 1$, while stars comprise around 10-20\%. The
shorter wavelength band
(9 and 11 $\mu$m) sources contain more stars than longer wavelength bands (15 and 18 $\mu$m) in MIR.

The color-color diagrams at near and mid infrared show broader distribution
in color-color space while stars are known to exist in narrow areas in these
diagrams. In particular, N2-N4 versus N3-N4 CC diagram is found to be
very useful in identifying stars as the stars have narrow ranges of these
colors.

The CC diagrams can also be used to identify galaxy types. For example,
AGN can be distinguished from relatively red N2-N4 color ($\gtsim 1.5$), while
starburst galaxies show very red $S7-S11$ colors ($>3$). The S7-S11 color appears
to be a good indicator for star formation rates.

We have shown the histograms in magnitude for S9W, S11, L15 and L18W bands. The
detailed shapes of the histograms depend on the pass band, but S9W and S11
have less steep slope than L15 and L18W at bright ends (brighter than 18 magnitude).
Such a difference have been caused by the contamination of stars for shorter wavelength bands, but
strong positive K-correction from the 7.7 and 11.3 $\mu$m may also be partially responsible.
The L15 and L18W band source distribution is consistent with that
of uniform sources in flat universe.
However, the detailed shape at faint part should be sensitive to
the luminosity and number density
evolution of star forming galaxies.


This work was supported in part by KRF grant No. 2006-341-
C00018. This work is based on observations with AKARI, a
JAXA project with the participation of ESA.
M.G.L and M.I were supported in part by a grant
(R01-2007-000-20336-0) from the Basic Research Program of the
Korea Science and Engineering Foundation.


\begin{thebibliography}{}

\bibitem[Bertin \& Arnouts (1996)] {bertin96}
Bertin, E., \& Arnouts, S. 1996, A\&AS, 117, 393
\bibitem[Gruppioniet al. (2002)] {gruppioni02}
Gruppioni C., Lari C., Pozzi F., Zamorani G., Franceschini A.,
Oliver S., Rowan-Robinson M., Serjeant S., 2002, MNRAS,
335, 831

\bibitem[Condon et al. 1996]{condon96}
Condon, J. J., Cotton, W. D., Greisen, E. W., Yin, Q. F., Perley, R. A., Taylor, G. B., \& Broderick, J. J.,
1998, AJ, 115, 1693

\bibitem[Hwang et al. (2007)]{hwang07}
Hwang, N., et al. \ 2007,  \apjs, R172, 583



\bibitem[Jenkner et al. (1990)]{jenkner90}
Jenkner, H., Lasker, B., Sturch, C., McLean, B., Shara, M., \& Russell, J.
\ 1990, \aj, 99, 2081

\bibitem[Lee et al. (2007)]{lee07}
Lee, H. M., et al. \ 2007, \pasj, 59, S529

\bibitem[Lorente et al. (2007)]{lorente07}
Lorente, R., Onaka, T., Ita, Y., Ohyama, Y., Tanabe, T., \& Pearson, C.
\ 2007, IRC User Data Manual, {\tt http://www.ir.isas.jaxa.jp/ASTRO-F/Observation/IDUM/IRC\_IDUM\_1.3.pdf}

\bibitem[Matsuhara et al. (2007)]{matsu07}
Matsuhara, H. et al.,\ 2007, \pasj, 59, S543

\bibitem[Matsuhara et al. (2006)]{matsu06}
Matsuhara, H. et al.,\ 2007, \pasj, 58, 673

\bibitem[Murakami et al. (2007)]{murakami07}
Murakami H. et al. \ 2007, \pasj, 59, S369

\bibitem[Onaka et al. (2007)]{onaka07}
Onaka, T. et al., \ 2007, \pasj, 59, S401

\bibitem[Silva et al. (1998)]{silva98}
Silva, L., Granato, G. L., Bressan, A., \& Danese, L. \ 1998, \apj, 509, 103

\bibitem[Takagi et al. (2007)]{takagi07}
Takagi, T., et al., \ 2007, \pasj, 59, S557

\bibitem[Wada et al. (2007)]{wada07}
Wada, K., et al., \ 2007, \pasj, 59, S515

\bibitem[Wada et al. (2007)]{wada07}
Wada, K., et al., \ 2008, \pasj, in press

\bibitem[van Dokkum (2001)]{dokkum01}
van Dokkum, P. G., \ 2001, \pasp, 113, 1420

\end{thebibliography}
\end{document}